\documentclass{sig-alternate-05-2015}

\usepackage{amsmath}
\usepackage{amssymb}
\usepackage{color}
\usepackage{graphicx}
\usepackage{algpseudocode}
\usepackage{algorithm}
\usepackage{array}
\usepackage{subfigure}
\usepackage{times}
\usepackage{balance}

\begin{document}

\CopyrightYear{2016}
\setcopyright{acmcopyright}
\conferenceinfo{KDD '16,}{August 13-17, 2016, San Francisco, CA, USA}
\isbn{978-1-4503-4232-2/16/08}\acmPrice{\$15.00}
\doi{http://dx.doi.org/10.1145/2939672.2939867}

\clubpenalty=10000
\widowpenalty = 10000
	
\setcopyright{acmcopyright}

\newtheorem{theorem}{Theorem}

\newcommand{\todo}[1]{\textcolor{red}{Todo: #1}}
\newcommand{\agp}[1]{\textcolor{red}{Aditya: #1}}
\newcommand{\system}{{\sc Squish}}

\newcommand{\probtree}{{\sc SquID}}
\newcommand{\probmodel}{{\sc SquID\-Model}}

\newcommand{\techreport}[1]{#1}
\newcommand{\paper}[1]{}

  \renewcommand{\sfdefault}{cmss}
  
\newcommand{\blue}[1]{\textcolor{blue}{#1}}

\newenvironment{denselist}{
    \begin{list}{\small{$\bullet$}}%
    {\setlength{\itemsep}{0ex} \setlength{\topsep}{0ex}
    \setlength{\parsep}{0pt} \setlength{\itemindent}{0pt}
    \setlength{\leftmargin}{1.5em}
    \setlength{\partopsep}{0pt}}}%
    {\end{list}}

\title{{\LARGE \system}: Near-Optimal Compression for Archival of Relational Datasets}

\numberofauthors{2}

\author{
\alignauthor
Yihan Gao\\
\affaddr{University of Illinois at Urbana-Champaign}\\
\affaddr{Urbana, Illinois}\\
\email{ygao34@illinois.com}
\alignauthor
Aditya Parameswaran\\
\affaddr{University of Illinois at Urbana-Champaign}\\
\affaddr{Urbana, Illinois}\\
\email{adityagp@illinois.com}
}

\maketitle
\begin{abstract}
Relational datasets are being generated at an alarmingly rapid rate across organizations and industries. Compressing these datasets could significantly reduce storage and archival costs. Traditional compression algorithms, e.g., gzip, are suboptimal for compressing relational datasets since they ignore the table structure and relationships between attributes. 

We study compression algorithms that leverage the relational structure to compress datasets to a much greater extent. We develop \system, a system that uses a combination of Bayesian Networks and Arithmetic Coding to capture multiple kinds of dependencies among attributes and achieve near-entropy compression rate. \system \\also supports user-defined attributes: users can instantiate new data types by simply implementing five functions for a new class interface. We prove the asymptotic optimality of our compression algorithm and conduct experiments to show the effectiveness of our system: \system~achieves a reduction of over 50\% in storage size relative to systems developed in prior work on a variety of real datasets.
\end{abstract}


\section{Introduction}

From social media interactions, commercial transactions, to scientific observations and the internet-of-things, 
relational datasets are being generated at an alarming rate. With these datasets, that are {\em either costly or impossible
to regenerate, there is a need for periodic archival,} for a variety of purposes, including long-term analysis or machine learning, 
historical or legal factors, or public or private access over a network. Thus, despite the declining costs of storage, 
compression of relational datasets is still important, and will stay important in the coming future.

One may wonder if compression of datasets is a solved problem. Indeed, there has been a variety of robust algorithms like Lempel-Ziv~\cite{ziv1977universal},
WAH~\cite{wu2004word}, and CTW~\cite{willems1995context} developed and used widely for data compression. However, these algorithms do not exploit the relational
structure of the datasets: attributes are often correlated or dependent on each other, and identifying and exploiting such correlations can lead to 
significant reductions in storage. In fact, there are many types of dependencies between attributes in a relational dataset.
For example, attributes could be functionally dependent on other attributes~\cite{babu2001spartan}, or the dataset could consist of several clusters of tuples, such that all the tuples within each cluster are similar to each other~\cite{jagadish2004itcompress}. The skewness of numerical attributes is another important source of redundancy that is overlooked by algorithms like Lempel-Ziv~\cite{ziv1977universal}: by designing encoding schemes based on the distribution of attributes, we can achieve much better compression rate than storing the attributes using binary/float number format.

There has been some limited work on compression of relational datasets, all in the recent past~\cite{babu2001spartan, davies1999bayesian, jagadish2004itcompress, raman2006wring}. In contrast with this line of prior work, \system~uses a combination of Bayesian Networks coupled with Arithmetic Coding~\cite{witten1987arithmetic}. 
Arithmetic Coding is a coding scheme designed for sequence of characters.
It requires an order among characters and probability distributions of characters conditioned on all preceding ones.
Incidentally, Bayesian Networks fulfill both requirements: 
the acyclic property of Bayesian Network provides us an order (i.e., the topological order), 
and the conditional probability distributions are also specified in the model. 
Therefore, Bayesian Networks and Arithmetic Coding are a perfect fit for relational dataset compression.

However, there are several challenges in using Bayesian Networks and Arithmetic Coding for compression. First, we need to identify a new objective function for learning a Bayesian Network, since conventional objectives like Bayesian Information Criterion~\cite{schwarz1978estimating} are not designed to minimize the size of the compressed dataset. Another challenge is to design a mechanism to support attributes with an infinite range (e.g., numerical and string attributes), since Arithmetic Coding assumes a finite alphabet for symbols, and therefore cannot be applied to those attributes. To be applicable to the wide variety of real-world datasets, it is essential to be able to handle numbers and strings.

We deal with these challenges in developing \system. As we show in this paper, the compression rate of \system~is near-optimal for all datasets that can be efficiently described using a Bayesian Network.
This theoretical optimality reflects in our experiments as well: {\em \system~achieves a reduction in storage on real datasets of over 50\% compared to the nearest competitor. } 
The reason behind this significant improvement is that most prior papers use sub-optimal techniques for compression.

In addition to being more effective at compression of relational datasets than prior work, \system~is also more {\em powerful}. To demonstrate that, 
we identify the following desiderata for a relational dataset compression system:
\begin{denselist}
	\item {\em Attribute Correlations (AC).} 
	A relational dataset compression system must be able to capture correlations across attributes.
	\item {\em Lossless and Lossy Compression (LC).} 
	A relational dataset compression system must be general enough to admit a user-specified error tolerance, 
	and be able to generate a compressed dataset in a lossy fashion, while respecting the error tolerance, further saving on storage.
	\item {\em Numerical Attributes (NA).} 
	In addition to string and categorical attributes, a relational dataset compression system must be able to capture numerical attributes, 
	that are especially	common in scientific datasets.
	\item {\em User-defined Attributes (UDA).} 
	A relational dataset compression system must be able to admit new types of attributes that do not fit into 
	either string, numerical, or categorical types.
\end{denselist}

In contrast to prior work~\cite{babu2001spartan, davies1999bayesian, jagadish2004itcompress, raman2006wring}---see Table~\ref{tbl_intro}---our system, \system, can capture all of these desiderata. To support UDA (User Defined Attributes),
\system~surfaces a new class interface, called the {\sc SquID} (short for \system~Interface for Data types): 
users can instantiate new data types by simply implementing the required five functions. 
This interface is remarkably powerful, especially for datasets in specialized domains.
For example, a new data type corresponding to genome sequence data can be implemented by a user using a few hundred lines of code. By encoding domain knowledge into the data type definition, we can achieve a significant higher compression rate than using ``universal'' compression algorithms like Lempel-Ziv~\cite{ziv1977universal}.

\begin{table}[h]
	\centering
	\vspace{-5pt}
	\begin{tabular}{c|c|c|c|c}
		System & AC & NA & LC & UDA \\
		\hline \hline
		\system & Y & Y & Y & Y \\
		\hline
		Spartan~\cite{babu2001spartan} & Y & Y & Y & N \\
		\hline
		ItCompress~\cite{jagadish2004itcompress} & Y & Y & Y & N \\
		\hline
		Davies and Moore~\cite{davies1999bayesian} & Y & N & N & N \\
		\hline
		Raman and Swart~\cite{raman2006wring} & N & N & N & N \\
	\end{tabular}
			\vspace{-5pt}
		\caption{Features of Our System contrasted with Prior Work}\label{tbl_intro}
		\vspace{-5pt}
\end{table}

The rest of this paper is organized as follows. In Section~\ref{sec_prelim}, we formally define the problem of relational dataset compression, and briefly explain the concepts of Arithmetic Coding. In Section~\ref{sec_algorithm} we discuss Bayesian Network learning and related issues, while details about Arithmetic Coding are discussed in Section~\ref{sec_implementation}. In Section~\ref{sec_optimality}, we \techreport{use examples to illustrate the effectiveness of \system~and }prove the asymptotic optimality of the compression algorithm. In Section~\ref{sec_experiment}, we conduct experiments to compare \system~with prior systems and evaluate its running time and parameter sensitivity. We describe related work in Section~\ref{sec_related_work}. All our proofs can be found \techreport{in the appendix}\paper{in our technical report~\cite{techreport}, along with a brief review of Bayesian Networks and illustrative examples of our compression algorithm}.

The source code of \system~is available on GitHub:  https://gith ub.com/Preparation-Publication-BD2K/db\_compress

\section{Preliminaries}\label{sec_prelim}
\noindent In this section, we define our problem more formally,
and provide some background on \techreport{Bayesian Networks
and }Arithmetic Coding.
\subsection{Problem Definition}\label{sec_prob_def}

We follow the same problem definition proposed by Babu et al.~\cite{babu2001spartan}.
Suppose our dataset consists of a single relational table $T$, with $n$ rows and $m$ columns (our techniques extend to multi-relational case as well). Each row of the table is referred to as a tuple and each column of the table is referred to as an attribute. We assume that each attribute has an associated domain that is known to us. For instance, this information could be described when the table schema is specified. 

The goal is to design a compression algorithm $A$ and a decompression algorithm $B$, such that $A$ takes $T$ as input and outputs a compressed file $C(T)$, and $B$ takes the compressed file $C(T)$ as input and outputs $T'$ as the approximate reconstruction of $T$. The goal is to minimize the file size of $C(T)$ while ensuring that the recovered $T'$ is close enough to $T$.


The closeness constraint of $T'$ to $T$ is defined as follows: For each numerical attribute $i$, for each tuple $t$ and the recovered tuple $t'$, $|t_i - t'_i| \leq \epsilon_i$, where $t_i$ and $t'_i$ are the values of attribute $i$ of tuple $t$ and $t'$ respectively, and $\epsilon_i$ is error threshold parameter provided by the user. For non-numerical attributes, the recovered attribute value must be exactly the same as the original one: $t_i = t'_i$.
Note that this definition subsumes lossless compression as a special case with $\epsilon_i = 0$. 

\techreport{
\subsection{Bayesian Network}\label{sec_bayesian}

Bayesian Networks~\cite{koller2009probabilistic} are widely used probabilistic models to describe the correlations between a collection of random variables. A Bayesian Network can be described using a directed acyclic graph among random variables, together with a set of conditional probability distributions for each random variable that only depends on the value of parent random variables in the graph (i.e., the neighbor nodes associated with incoming links). 

More formally, a Bayesian Network $\mathcal{B} = (\mathcal{G}, (\mathcal{M}_1, \ldots, \mathcal{M}_n))$ consists of two components: The structure graph $\mathcal{G} = (V, E)$ is a directed acyclic graph with $n$ vertices $v_1, \ldots, v_n$ corresponding to $n$ random variables (denote them as $X_1, \ldots, X_n$). For every random variable $X_i$, $\mathcal{M}_i$ is a model describing the conditional distribution of $X_i$ conditioned on $\mathbf{X}_{parent(i)} = \{X_j : j \in parent(i)\}$, where $parent(i) = \{j :  (v_j, v_i) \in E\}$ is called the set of parent nodes of $v_i$ in $\mathcal{G}$. 

Figure~\ref{fig_bn_example} shows an example Bayesian Network with three random variables $X_1, X_2, X_3$. The structure graph $\mathcal{G}$ is shown at top of the figure and the bottom side shows the models $\mathcal{M}_1, \mathcal{M}_2, \mathcal{M}_3$.

\begin{figure}[h]
	\centering
	\includegraphics[width = 7.5cm]{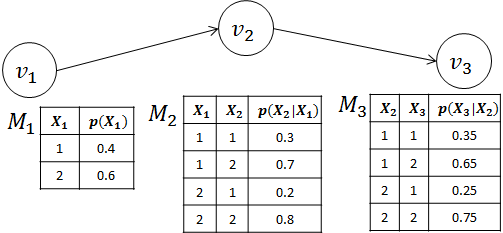}
	\caption{Bayesian Network Example}\label{fig_bn_example}
\end{figure}

In this work, we will use Bayesian Networks to model the correlations between attributes within each tuple. In particular, the random variables will correspond to attributes in the dataset and edges will capture correlations. Thus, each tuple can be viewed as a joint-instantiation of values of the random variables in the Bayesian Network, or equivalently, a sample from the probability distribution described by the Bayesian Network.
}

\begin{figure*}[ht]
	\centering
	\vspace{-5pt}
	\includegraphics[width = 16cm]{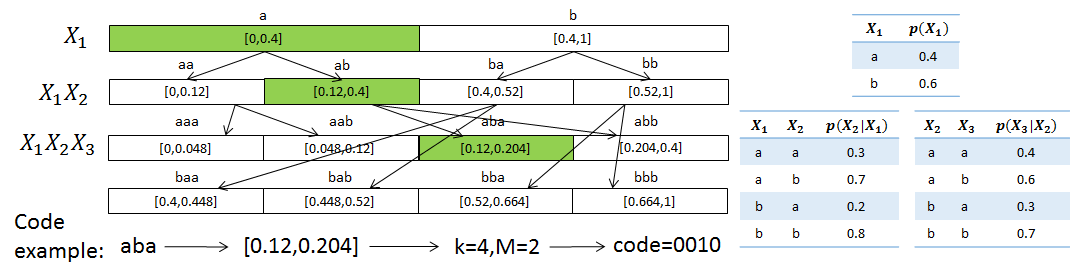}
	\vspace{-10pt}
	\caption{Arithmetic Coding Example}\label{fig_arithmetic_coding}
	\vspace{-10pt}
\end{figure*} 

\subsection{Arithmetic Coding}\label{sec_arithmetic}

Arithmetic coding~\cite{witten1987arithmetic,mackay2003information} is a state-of-the-art adaptive compression algorithm for a sequence of dependent characters. Arithmetic coding assumes as a given a conditional probability distribution model for any character, conditioned on all preceding characters. If the sequence of characters are indeed generated from the probabilistic model, then arithmetic coding can achieve a near-entropy compression rate~\cite{mackay2003information}.

Formally, arithmetic coding is defined by a finite ordered alphabet $\mathcal{A}$, and a probabilistic model for a sequence of characters that specifies the probability distribution of each character $X_k$ conditioned on all precedent characters $X_1, \ldots, X_{k-1}$. Let $\{a_n\}$ be any string of length $n$. To compute the encoded string for $\{a_n\}$, we first compute a probability interval for each character $a_k$:
\begin{align*} 
[l_k, r_k] = & [p(X_k < a_k|X_1 = a_1, \ldots, X_{k-1} = a_{k-1}), \\
			 & p(X_k \leq a_k|X_1 = a_1, \ldots, X_{k-1} = a_{k-1})]
\end{align*}

We define the product of two probability interval as:
$$[l_1, r_1] \circ [l_2, r_2] = [l_1 + (r_1 - l_1) l_2, l_1 + (r_1 - l_1) r_2]$$

The probability interval for string $\{a_n\}$ is the product of probability intervals of all the characters in the string:
$$ [l,r] = [l_1,r_1] \circ [l_2,r_2] \circ \ldots \circ [l_n, r_n] $$

Let $k$ be the smallest integer such that there exists a non-negative integer $0 \leq M < 2^k$ satisfying:
$$ l \leq 2^{-k}M, r \geq 2^{-k} (M + 1) $$
Then the $k$-bit binary representation of $M$ is the encoded bit string of $\{a_n\}$.

An example to illustrate how arithmetic coding works can be found in Figure~\ref{fig_arithmetic_coding}. The three tables at the right hand side specify the probability distribution of the string $a_1 a_2 a_3$. The blocks at the left hand show the associated probability intervals for the strings: for example, ``aba'' corresponds to $[0.12, 0.204] = [0, 0.4] \circ [0.3, 1] \circ [0, 0.3]$. As we can see, the intuition of arithmetic aoding is to map each possible string $\{a_n\}$ to disjoint probability intervals. By using these probability intervals to construct encoded strings, we can make sure that no code word is a prefix of another code word.


Notice that the length of the product of probability intervals is exactly the product of their lengths. Therefore, the length of each probability interval is exactly the same as the probability of the corresponding string $\mathbf{Pr}(\{a_n\})$. Using this result, the length of encoded string can be bounded as follows:
$$ \textit{len}(\textit{binary\_code}(\{a_n\})) \leq \lfloor - \log_2 \mathbf{Pr}(\{a_n\}) \rfloor + 2 $$

\section{Structure Learning}\label{sec_algorithm}

\begin{figure}[h]
    \centering
    \includegraphics[width = 6cm]{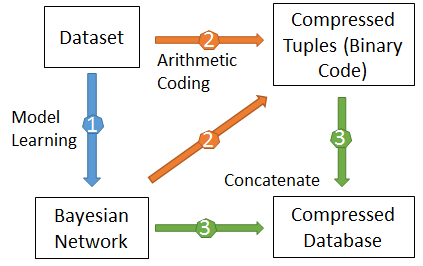}
    \vspace{-10pt}
    \caption{Workflow of the Compression and Decompression Algorithm}\label{fig_workflow_overview}
    \vspace{-15pt}
\end{figure}

The overall workflow of \system\ is illustrated in Figure~\ref{fig_workflow_overview}. 
\system \\ uses a combination of Bayesian networks and arithmetic coding for compression.
The workflow of the compression algorithm is the following:

\begin{enumerate}
	\item
	Learn a Bayesian network structure from the dataset, which captures the dependencies between attributes in the structure graph, and models the conditional probability distribution of each attribute conditioned on all the parent attributes.
	\item
	Apply arithmetic coding to compress the dataset, using the Bayesian network as probabilistic models.
	\item
	Concatenate the model description file (describing the Bayesian network model) and compressed dataset file.
\end{enumerate}
In this section, we focus on the first step of this workflow.
We focus on the remaining steps (along with decompression) in Section~\ref{sec_implementation}.


Although the problem of Bayesian network learning has been extensively studied in literature~\cite{koller2009probabilistic}, conventional objectives like Bayesian Information Criterion (BIC)~\cite{schwarz1978estimating} are suboptimal for the purpose of compressing datasets. In Section~\ref{sec_model_learning}, we derive the correct objective function for learning a Bayesian network that minimizes the size of the compressed dataset and explain how to modify existing Bayesian network learning algorithms to optimize this objective function.

The general idea about how to apply arithmetic coding on a Bayesian network is straightforward: since the graph encoding the structure of a Bayesian Network is acyclic, we can use any topological order of attributes and treat the attribute values as the sequence of symbols in arithmetic coding. However, arithmetic coding does not naturally apply to non-categorical attributes.
In Section~\ref{sec_probtree}, we introduce \probtree, the mechanism for supporting non-categorical and arbitrary user-defined attribute types in \system. \probtree~is the interface for every attribute type in \system, and example \probtree s for categorical, numerical and string attributes are demonstrated in Section~\ref{sec_example_model} to illustrate the wide applicability of this interface. We describe the \probtree \ API in Section~\ref{sec_api}.



\subsection{Learning a Bayesian Network: The Basics}\label{sec_model_learning}

Many Bayesian network learning algorithms search for the optimal Bayesian network by minimizing some objective function (e.g., negative log-likelihood, BIC~\cite{schwarz1978estimating}). These algorithms usually have two separate components~\cite{koller2009probabilistic}:
\begin{denselist}
	\item
	A combinatorial optimization component that searches for a graph with optimal structure.
	\item
	A score evaluation component that evaluates the objective function given a graph structure.
\end{denselist}

The two components above are independent in many algorithms. In that case, we can modify an existing Bayesian network learning algorithm by changing the score evaluation component, while still using the same combinatorial optimization component. In other words, for any objective function, as long as we can efficiently evaluate it based on a fixed graph structure, we can modify existing Bayesian network learning algorithms to optimize it. 

In this section, we derive a new objective function for learning a Bayesian network that minimizes the size of compressed dataset. We show that the new objective function can be evaluated efficiently given the structure graph. Therefore existing Bayesian Network learning algorithms can be used to optimize it.


Suppose our dataset $D$ consists of $n$ tuples, and each tuple $t_i$ contains $m$ attributes $a_{i1}, a_{i2}, \ldots, a_{im}$. Let $\mathcal{B}$ be a Bayesian network that describes a joint probability distribution over the attributes. Clearly, $\mathcal{B}$ contains $m$ nodes, each corresponding to an attribute.

The total description length of $D$ using $\mathcal{B}$ is $\textbf{S}(D|\mathcal{B}) = \textbf{S}(\mathcal{B}) + \textbf{S}(\textit{Tuples}|\mathcal{B})$,
where $\textbf{S}(\mathcal{B})$ is the size of description file of $\mathcal{B}$, and $\textbf{S}(\textit{Tuples}|\mathcal{B})$ is the total length of encoded binary strings of tuples using arithmetic coding. For the model description length $\textbf{S}(\mathcal{B})$, we have $\textbf{S}(\mathcal{B}) = \sum_{i = 1}^m \textbf{S}(\mathcal{M}_i)$, 
where $m$ is the number of attributes in our dataset, and $\mathcal{M}_1, \ldots, \mathcal{M}_m$ are the models for each attribute in $\mathcal{B}$. The expression
$\textbf{S}(\textit{Tuples}|\mathcal{B})$ is just the sum of the $\textbf{S}(t_i|\mathcal{B})$s (the lengthes of the encoded binary string for each $t_i$). 

We have the following decomposition of $\textbf{S}(t_i|\mathcal{B})$:
\begin{align*}
\textbf{S}(t_i|\mathcal{B}) \approx & - \sum_{j=1}^m \log_2 \mathbf{Pr}(a_{ij}|\textit{parent}(a_{ij}), \mathcal{M}_j) \\
									& - \sum_{j=1}^m \textit{num}(a_{ij}) \log_2 \epsilon_j + \text{const}
\end{align*}
where $\textit{parent}(a_{ij})$ is the set of parent attributes of $a_{ij}$ in $\mathcal{B}$, $\textit{num}(a)$ is the indicator function of whether $a$ is a numerical attribute or not, and $\epsilon_j$ is the maximum tolerable error for attribute $a_{ij}$. We will justify this decomposition in Section~\ref{sec_example_model}, after we introduce the encoding scheme for numerical attributes.


Therefore, the total description length $\textbf{S}(D|\mathcal{B})$ can be decomposed as follows:
\begin{align*}
\textbf{S}(D|\mathcal{B}) \approx & \sum_{j=1}^m [\textbf{S}(\mathcal{M}_j) - \sum_{t_i \in \textit{DB}} \log_2 \mathbf{Pr}(a_{ij}|\textit{parent}(a_{ij}), \mathcal{M}_j)] \\
				& - n (\sum_{j=1}^m \textit{num}(a_j) \log_2 \epsilon_j + \text{const})
\end{align*}

Note that the term in the second line does not depend on either $\mathcal{B}$ or $\mathcal{M}_i$. Therefore we only need to optimize the first summation. We denote each term in the first summation on the right hand side as $obj_j$:
$$ obj_j = \textbf{S}(\mathcal{M}_j) - \sum_{t_i \in \textit{DB}} \log \mathbf{Pr}(a_{ij}|\textit{parent}(a_{ij}), \mathcal{M}_j) $$

For each $obj_j$, if the network structure (i.e., $\textit{parent}(a_{ij})$) is fixed, then $obj_j$ only depends on $\mathcal{M}_j$. In that case, optimizing $\textbf{S}(D|\mathcal{B})$ is equivalent to optimizing each $obj_j$ individually. In other words, if we fix the Bayesian network structure in advance, then the parameters of each model can be learned separately. 

Optimizing $obj_j$ on $\mathcal{M}_j$ is exactly the same as maximizing likelihood. For many models, a closed-form solution for identifying maximum likelihood parameters exists. In such cases, the optimal $\mathcal{M}_j$ can be quickly determined and the objective function $\textbf{S}(D|\mathcal{B})$ can be computed efficiently.

\vspace{2mm}

\noindent \textbf{Structure Learning}

\vspace{1mm}

\noindent In general, searching for the optimal Bayesian network structure is NP-hard~\cite{koller2009probabilistic}. In \system, we implemented a simple greedy algorithm for this task. The algorithm starts with an empty seed set, and repeatedly finds new attributes with the lowest $obj_j$, and adds these new attributes to the seed set. \techreport{The pseudo-code is shown in Algorithm~\ref{alg_structure_learning}.}\paper{The pseudo-code can be found in our technical report~\cite{techreport}.}

\techreport{
\begin{algorithm}
	\caption{Greedy Structure Learning Algorithm}
	\label{alg_structure_learning}
	\begin{algorithmic}
		\Function{LearnStructure}{}
		\State $\text{seed} \leftarrow \emptyset$
		\For{$i = 1$ to $m$}
			\For{$j = 1$ to $m$}				
				\State $\text{parent} \leftarrow \emptyset$
				\While{true}
					\State $\text{best\_model\_score} \leftarrow obj(\text{parent} \rightarrow j)$
					\State $\text{best\_model} \leftarrow \text{parent}$
					\For{$k \in \text{seed}$}
						\State $t \leftarrow obj(\text{parent} \cup \{k\} \rightarrow j)$
						\If{$t < \text{best\_model\_score}$}
							\State $\text{best\_model\_score} \leftarrow t$
							\State $\text{best\_model} \leftarrow \text{parent} \cup \{k\}$
						\EndIf
					\EndFor
					\If{$\text{best\_model} = \text{parent}$}
						\State \textbf{break}
					\Else
						\State $\text{parent} \leftarrow \text{best\_model}$
					\EndIf
				\EndWhile
				\State $\text{score}_j = obj(\text{parent} \rightarrow j)$
			\EndFor
			\State Find $j$ with minimum $\text{score}_j$
			\State $\text{seed} \leftarrow \text{seed} \cup \{j\}$
		\EndFor
		\EndFunction
	\end{algorithmic}
\end{algorithm}
}

\paper{The greedy algorithm} \techreport{Algorithm~\ref{alg_structure_learning}} has a worst case time complexity of $O(m^4 n)$ where $m$ is the number of columns and $n$ is the number of tuples in the dataset. For large datasets, even this simple greedy algorithm is not fast enough. However, note that the objective values \techreport{$obj(\text{parent} \rightarrow j)$}\paper{$obj_j$} are only used to compare different models. So we do not require exact values for them, and some rough estimation would be sufficient. Therefore, we can use only a subset of data for the structure learning to improve efficiency. 

\subsection{Supporting Complex Attributes}\label{sec_probtree}

\vspace{2mm}

\noindent \textbf{Encoding and Decoding Complex Attributes}

\vspace{1mm}

\noindent Before applying arithmetic coding on a Bayesian network to compress the dataset as we stated earlier, there are two issues that we need to address first:
\begin{itemize}
	\item
	Arithmetic Coding requires a finite alphabet for each symbol. However, it is natural for attributes in a dataset to have infinite range (e.g., numerical attributes, strings). 
	\item
	In order to support user-defined data types, we need to allow the user to specify a probability distribution over an unknown data type.
\end{itemize}

To address these difficulties, we introduce the concept of {\probtree}, short for \system~Interface for Data types. A \probtree~is a (possibly infinite) decision tree~\cite{witten2005data} with non-negative probabilities associated with edges in the tree, such that for every node $v$, the probabilities of all the edges connecting $v$ and $v$'s children sum to one. 

\begin{figure}[h]
	\centering
	\vspace{-10pt}
	\includegraphics[width = 6cm]{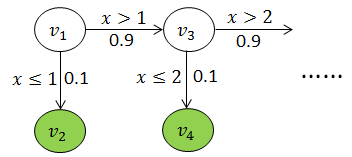}
	\vspace{-10pt}
	\caption{\probtree~Example}\label{fig_prob_tree_example}
	\vspace{-10pt}
\end{figure}

Figure~\ref{fig_prob_tree_example} shows an example infinite \probtree~for a positive numerical attribute $X$. As we can see, each edge is associated with a decision rule and a non-negative probability. For each non-leaf node $v_{2k-1}$, the decision rules on the edges between $v_{2k-1}$ and its children $v_{2k}$ and $v_{2k+1}$ are $x \leq k$ and $x > k$ respectively. Note that these two decision rules do not overlap with each other and covers all the possibilities. The probabilities associated with these two rules sum to $1$, which is required in \probtree. This \probtree~describes the following probability distribution over $X$:
$$ \mathbf{Pr}(X \in (k - 1, k]) = 0.9^{k-1} \times 0.1 $$

In Section~\ref{sec_implementation}, we will show that we can encode or decode an attribute using Arithmetic Coding if the probability distribution of this attribute can be represented by a \probtree.

As shown in Figure~\ref{fig_prob_tree_example}, a \probtree~naturally controls the maximum tolerable error in a lossy compression setting. Each leaf node $v$ corresponds to a subset $A_v$ of attribute values such that for every $a \in A_v$, if we start from the root and traverse down according to the decision rules, we will eventually reach $v$. As an example, in Figure~\ref{fig_prob_tree_example}, for each leaf node $v_{2k}$ we have $A_{v_{2k}} = (k-1,k]$. Let $a_v$ be the representative attribute value of a leaf node $v$, then the maximum possible recovery error for $v$ is:
$$\epsilon_v = \sup_{a \in A_v} \text{distance}(a, a_v)$$

Let $T_i$ be the \probtree~corresponding to the $i$th attribute. As long as for every $v \in T_i$, $\epsilon_v$ is less than or equal to the maximum tolerable error $\epsilon_i$, we can satisfy the closeness constraint (defined in Section~\ref{sec_prob_def}).

\vspace{2mm}

\noindent \textbf{Using User-defined Attributes as Predictors}

\vspace{2mm}

To allow user-defined attributes to be used as predictors for other attributes, we introduce the concept of attribute interpreters, which translate attributes into either categorical or numerical values. In this way, these attributes can be used as predictors for other attributes. 

The attribute interpreters can also be used to capture the essential features of an attribute. For example, a numerical attribute could have a categorical interpreter that better captures the internal meaning of the attribute. This process is similar to the feature extraction procedure in many data mining applications, and may improve compression rate.

\subsection{Example {\large \probtree s}}\label{sec_example_model}

In \system, we have implemented models for three primitive data types. We intended these models to both illustrate how \probtree s can be used to define probability distributions, and also to cover most of the basic data types, so the system can be used directly by casual users without writing any code. 

We implemented models for the following types of attributes:
\begin{denselist}
	\item
	Categorical attributes with finite range.
	\item
	Numerical attributes, either integer or float number.
	\item
	String attributes
\end{denselist}

\techreport{In general, an attribute can be represented by \probtree~if it can be constructed using several simple steps. For any given probability distribution, there might be more than one possible \probtree~design. As long as the underlying distribution is correctly encoded, the compression rate will be the same. However, the size of the \probtree could vary and will directly affect the efficiency of the compression algorithm, so \probtree should be carefully designed if high efficiency is desired.}

\vspace{2mm}

\noindent \textbf{Categorical Attributes}

\vspace{2mm}

The distribution over a categorical attributes can be represented using a trivial one-depth \probtree. 

\vspace{2mm}

\noindent \textbf{Numerical Attributes}

\vspace{2mm}

\begin{figure}[h]
	\centering
	\includegraphics[width = 7cm]{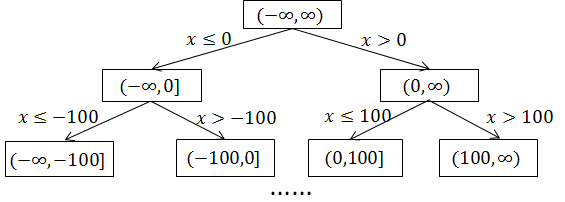}
	\vspace{-15pt}
	\caption{\probtree~for numerical attributes}\label{fig_numerical_prob_tree}
	\vspace{-10pt}
\end{figure}

For a numerical attribute, we construct the \probtree~using the idea of {\em bisection}. Each node $v$ is marked with an upper bound $v_r$ and a lower bound $v_l$, so that every attribute value in range $(v_l,v_r]$ will pass by $v$ on its path from the root to the corresponding leaf node. Each node has two children and a bisecting point $v_m$, such that the two children have ranges $(v_l, v_m]$ and $(v_m, v_r]$ respectively. The branching process stops when the range of the interval is less than $2 \epsilon$, where $\epsilon$ is the maximum tolerable error.
Figure~\ref{fig_numerical_prob_tree} shows an example \probtree~for numerical attributes.

Since each node represents a continuous interval, we can compute its probability using the cumulative distribution function. The branching probability of each node is:
\begin{align*}
 & (\mathbf{Pr}(\textit{left branch}), \mathbf{Pr}(\textit{right branch})) \\
= & (\frac{\mathbf{Pr}(v_l < X \leq v_m)}{\mathbf{Pr}(v_l < X \leq v_r)}, \frac{\mathbf{Pr}(v_m < X \leq v_r)}{\mathbf{Pr}(v_l < X \leq v_r)})
\end{align*}


Clearly, the average number of bits that is needed to encode a numerical attribute depends on both the probability distribution of the attribute and the maximum tolerable error $\epsilon$. The following theorem gives us a lower bound on the average number of bits that is necessary for encoding a numerical attribute (the proof can be found in \techreport{the appendix}\paper{our technical report~\cite{techreport}}):

\begin{theorem}\label{thm_numerical}
	Let $X \in \mathcal{X} \subseteq \mathbb{R}$ be a numerical random variable with continuous support $\mathcal{X}$ and probability density function $f(X)$. Let $g: \mathcal{X} \rightarrow \{0,1\}^*$ be any uniquely decodable encoding function, and $h: \{0,1\}^* \rightarrow \mathcal{X}$ be any decoding function. If there exists a function $\rho:\mathcal{X} \rightarrow \mathbb{R}^+$ such that:
	\begin{equation}\label{eqn_self_bounded}
	\forall x, y \in \mathcal{X}, |x - y| < 2\epsilon \Rightarrow |f(x) - f(y)| \leq \rho(x) f(x) |x - y|
	\end{equation}
	and $g, h$ satisfies the $\epsilon$-closeness constraint:
	$$ \forall x \in \mathcal{X}, |h(g(x)) - x| \leq \epsilon $$
	Then
	\begin{align*}
	E_X[len(g(X))] \geq & E_X[- \log_2 f(X)] - \\
						& E_X[\log_2(2 \epsilon \rho(X) + 1)] - \log_2 \epsilon - 2
	\end{align*}
	Furthermore, if $g$ is the bisecting code described above, then
	\begin{align*}
	E_X[len(g(X))] \leq & E_X[- \log_2 f(X)] - \log_2 l + \\
						& E_X[\max(\log_2 \rho(X) + \log_2 l, 0)] + 4
	\end{align*}

	where $l = \min_{v} (v_r - v_l)$ is the minimum length of probability intervals in the tree.
\end{theorem}

Equation~(\ref{eqn_self_bounded}) is a mild assumption that holds for many common probability distributions, including uniform distribution, Gaussian distribution, and Laplace distribution~\cite{abramowitz1964handbook}. 

To understand the intuition behind the results in Theorem~\ref{thm_numerical}. Let us consider a Gaussian distribution as an example:
$$ f(x|\mu,\sigma) = \frac{1}{\sigma \sqrt{2 \pi}} e^{-\frac{(x - \mu)^2}{2\sigma^2}} $$
In this case,
$$ \rho(x) = \frac{|x - \mu| + 2\epsilon}{\sigma^2} $$
Substituting into the first expression, we have:
\begin{align*}
E_X[len(g(X))] \geq & \log_2 \sigma - \log_2 \epsilon + \log_2 \sqrt{2 \pi} - \frac{3}{2} - \\
					& E_X[\log_2 (\frac{2 \epsilon(|X - \mu| + 2\epsilon)}{\sigma^2} + 1)]
\end{align*}
Note that when $\epsilon$ is small compared to $\sigma$ (e.g., $\epsilon < \frac{1}{10} \sigma$), the last term is approximately zero. Therefore, the number of bits needed to compress $X$ is approximately $\log_2 \frac{\sigma}{\epsilon}$.

Now consider the second result,
\begin{align*}
E_X[len(g(X))] \leq & \log_2 \sigma - \log_2 \frac{l}{2} + \log_2 \sqrt{2 \pi} + \frac{7}{2} + \\
& E_X[\max(\log_2 \frac{l(|X - \mu| + 2\epsilon)}{\sigma^2}, 0)]
\end{align*}
Let $l = 2 \epsilon$, and when $\epsilon < \frac{1}{10} \sigma$, the last term is approximately zero. Comparing the two results, we can see that the bisecting scheme achieves near optimal compression rate.

Theorem~\ref{thm_numerical} can be used to justify the decomposition in Section~\ref{sec_model_learning}. Recall that we used the following expression as an approximation of $len(g(X))$:
$$ len(g(X)) \approx - \log_2 f(X) - \log_2 \epsilon + \text{const} $$
Compared to either the upper bound or lower bound in Theorem~\ref{thm_numerical}, the only term we omitted is the term related to $\rho(X)$. As we have seen in the Gaussian case, this term is approximately zero when $\epsilon$ is smaller than $\frac{\sigma}{10}$ where $\sigma$ is the standard deviation parameter. The same conclusion is also true for the Laplace distribution~\cite{abramowitz1964handbook} and the uniform distribution.

\vspace{2mm}
\noindent \textbf{String Attributes}
\vspace{2mm}

The \probtree~for string attributes can be viewed as having two steps:
\begin{enumerate}
	\item
	determine the length of the string
	\item
	determine the characters of the string
\end{enumerate}
The length of a string can be viewed as an integer, so we can use exactly the same bisecting rules as for numerical attributes. After that, we use $n$ more steps to determine each character of the string, where $n$ is the string's length. The probability distribution of each character can be specified by conventional probabilistic models like the $k$-gram model.

\begin{table*}[t]
	\caption{Functions need to be implemented for \probtree~Template}\label{tbl_probtree}
	\begin{center}
		\begin{tabular}{| c | c |}
			\hline
			\textbf{Function} & \textbf{Description} \\
			\hline
			IsEnd & Return whether the current node is a leaf node. \\
			\hline
			GenerateBranch & Return the number of branches and the probability distribution associated with them. \\
			\hline
			GetBranch & Given an attribute value, return which branch does the value belong to. \\
			\hline
			ChooseBranch & Set the current node to another node at next level according to the given branch index. \\
			\hline
			GetResult & If the current node is a leaf node, return the representative attribute value of this node. \\
			\hline
		\end{tabular}
	\end{center}
\paper{	\vspace{-20pt} }
\end{table*}

\techreport{
\begin{table*}[t]
	\vspace{-25pt}
	\caption{Functions for \probmodel~Abstract Class}\label{tbl_probmodel}
	\begin{center}
		\begin{tabular}{| c | c |}
			\hline
			\textbf{Function} & \textbf{Description} \\
			\hline
			GetProbTree & Given the values of parent attributes, return a \probtree~instance. \\
			\hline
			ReadTuple & Read a tuple into the model. \\
			\hline
			EndOfData & Indicate the end of dataset. All gathered statistics can be used to find the optimal parameters. \\
			\hline
			GetModelCost & Return an estimate of the objective value $\textit{obj}_j = \textbf{S}(\mathcal{M}_j) - \sum_{t_i \in \textit{DB}} \log \mathbf{Pr}(a_{ij}|\textit{parent}(a_{ij}), \mathcal{M}_j)$. \\
			\hline
			WriteModel & Write the description of the model to a file. This function will be called only after EndOfData is called.\\
			\hline
			ReadModel & Read the description of the model from a file. This function will be called to initialize the \probmodel.\\
			\hline
		\end{tabular}
	\end{center}
	\vspace{-15pt}
\end{table*}
}
\subsection{{\large \probtree}~API}\label{sec_api}

In \system, \probtree~is defined as an abstract class~\cite{cppreference}. There are five functions that are required to be implemented in order to define a new data type using \probtree. These five functions allow the system to interactively explore the \probtree~class: initially, the current node pointer is set to the root of \probtree; each function will either acquire information about the current node, or move the current node pointer to one of its children. Table~\ref{tbl_probtree} lists the five functions together with their high level description.

We also develop another abstract class called \probmodel. A \probmodel~first reads in all the tuples in the dataset, then generates a \probtree~instance and an estimation of the objective value $\textit{obj}_j$ derived in Section~\ref{sec_model_learning}:
$$ \textit{obj}_j = \textbf{S}(\mathcal{M}_j) - \sum_{t_i \in \textit{DB}} \log \mathbf{Pr}(a_{ij}|\textit{parent}(a_{ij}), \mathcal{M}_j) $$

There are two reasons behind this design:
\begin{itemize}
	\item
	For a parametric \probtree, the parameters need to be learned using the dataset at hand.
	\item
	The Bayesian network learning algorithm requires an estimation of the objective value. Although it is possible to compute the objective value by actually compressing the attributes, in many cases it is much more efficient to approximate it directly.
\end{itemize}

\probmodel~requires six functions to be implemented. These functions allow the \probmodel~to iterate over the dataset and generate \probtree~instances. \paper{The specification of these functions and the psuedo-code of their interactions with \probtree~can be found in our technical report~\cite{techreport}.}\techreport{Table~\ref{tbl_probmodel} listed the six functions together with their high level explanation.}

A \probmodel~instance is initialized with the target attribute and the set of predictor attributes. After that, the model instance will read over all tuples in the dataset and need to decide the optimal choice of parameters. The model also needs to return an estimate of the objective value $obj_j$, which will be used in the Bayesian network structure learning algorithm \techreport{(i.e., Algorithm~\ref{alg_structure_learning})} to compare models. Finally, \probmodel~should be able to generate \probtree~instances based on parent attribute values.

\techreport{
\vspace{2mm}

\noindent \textbf{Usage of Abstract Classes in the System}

\vspace{2mm}

Algorithm~\ref{alg_api_calls} shows how our system interacts with these abstract classes. In the model learning phase, the \probmodel~is used to estimate the objective value $obj_j$. In the compression and decompression phase, \probmodel~is used to generate \probtree~instances, which are then used to generate probability intervals for tuples.

\begin{algorithm}
	\caption{API Calls to Abstract Classes}
	\label{alg_api_calls}
	\begin{algorithmic}
		\Function{ComputeObj}{$\text{parent} \rightarrow j$}
		\State $\text{model} \leftarrow \textbf{new } \text{\probmodel}(\text{parent} \rightarrow j)$
		\For{$i = 1$ to $n$}
		\State $\text{model}.ReadTuple(t_i)$
		\EndFor
		\State $\text{model}.EndOfData()$
		\State \Return $\text{model}.GetModelCost()$
		\EndFunction
		
		\Function{Compression}{$t_i, \{M_1, \ldots, M_m\}$}
		\State $\text{PIseq} \leftarrow \emptyset$
		\For{$j = 1$ to $m$}
		\State $T_j \leftarrow M_j.GetProbTree(t_i, \text{parent}_j)$
		\While{\textbf{not} $T_j.IsEnd()$}
		\State $branches \leftarrow T_j.GenerateBranch()$
		\State $b \leftarrow T_j.GetBranch(a_{ij})$
		\State $[l,r] \leftarrow \text{ComputeProbabilityInterval}(branches, b)$
		\State $\text{PIseq} \leftarrow \text{PIseq} + \{[l,r]\}$
		\State $T_j.ChooseBranch(b)$
		\EndWhile
		\State $a_{ij} \leftarrow T_j.GetResult()$
		\EndFor
		\State \Return $\text{ArithmeticCoding}(\text{PIseq})$
		\EndFunction
		
		\Function{Decompression}{$\{M_1, \ldots, M_m\}$}
		\State $t_i \leftarrow \emptyset$
		\State $d \leftarrow \textbf{new} \text{ Decoder}()$
		\For{$j = 1$ to $m$}
		\State $T_j \leftarrow M_j.GetProbTree(t_i, \text{parent}_j)$
		\While{\textbf{not} $T_j.IsEnd()$}
		\State $branches \leftarrow T_j.GenerateBranch()$
		\State $b \leftarrow d.GetNextBranch(branches)$
		\State $T_j.ChooseNextBranch(b)$
		\EndWhile
		\State $a_{ij} \leftarrow T_j.GetResult()$
		\EndFor
		\State \Return $t_i$
		\EndFunction
	\end{algorithmic}
\end{algorithm}
}

\section{Compression and Decompression}\label{sec_implementation}

In this section, we discuss how we can use arithmetic coding correctly for compression and decompression given a Bayesian Network. In Section~\ref{sec_compression}, we discuss implementation details that ensure the correctness of arithmetic coding using finite precision float numbers. \techreport{Section~\ref{sec_delta} talks about an algorithm that we borrow from prior work for additional compression. }In Section~\ref{sec_decompression} we describe the decompression algorithm. 

\subsection{Compression}\label{sec_compression}

We use the same notation as in Section~\ref{sec_model_learning}: a tuple $t$ contains $m$ attributes, and without loss of generality we assume that they follow the topological order of the Bayesian network:
$$ t = \{a_{1}, \ldots, a_{m}\}, \textit{parent}(a_{j}) \subseteq \{a_{1}, \ldots, a_{j-1}\} $$

We first compute a probability interval for each branch in a \probtree. For each \probtree~$T$, we define $\text{PI}_T$ as a mapping from branches of $T$ to probability intervals. The definition is similar to the one in Section~\ref{sec_arithmetic}: let $v$ be any non-leaf node in $T$, suppose $v$ has $k$ children $u_1, \ldots, u_k$, and the edge between $v$ and $u_i$ is associated with probability $p_i$, then $\text{PI}_T(v \rightarrow u_i)$ is defined as:
$$ \text{PI}(v \rightarrow u_i) = [\sum_{j < i} p_j, \sum_{j \leq i} p_j] $$

Now we can compute the probability interval of $t$. Let $T_j$ be the \probtree~for $a_j$ conditioned on its parent attributes. Denote the leaf node in $T_j$ that $a_j$ corresponds to as $v_j$. Suppose the path from the root of $T_j$ to $v_j$ is $u_{j1} \rightarrow u_{j2} \rightarrow \ldots \rightarrow u_{jk_j} \rightarrow v_j$. Then, the probability interval of tuple $t$ is:
\begin{align*}
[L,R] = & \text{PI}_{T_1}(u_{11} \rightarrow u_{12}) \circ \ldots \circ \text{PI}_{T_1}(u_{1k_1} \rightarrow v_1) \circ \\
	    & \text{PI}_{T_2}(u_{21} \rightarrow u_{22}) \circ \ldots \circ \text{PI}_{T_2}(u_{2k_2} \rightarrow v_2) \circ \ldots \circ \\
	    & \text{PI}_{T_m}(u_{m1} \rightarrow u_{m2}) \circ \ldots \circ \text{PI}_{T_m}(u_{mk_m} \rightarrow v_m)
\end{align*}
where $\circ$ is the probability interval multiplication operator defined in Section~\ref{sec_arithmetic}. The code string of tuple $\mathbf{t}$ corresponds to the largest subinterval of $[L,R]$ of the form $[2^{-k}M, 2^{-k}(M+1)]$ as described in Section~\ref{sec_arithmetic}.

In practice, we cannot directly compute the final probability interval of a tuple: there could be hundreds of probability intervals in the product, so the result can easily exceed the precision limit of a floating-point number. 

Algorithm~\ref{alg_encode} shows the pseudo-code of the precision-aware compression algorithm. We leverage two tricks to deal with the finite precision problem: the classic early bits emission trick~\cite{langdon1984introduction} is described in Section~\ref{sec_early_bit_emission}; the new deterministic approximation trick is described in Section~\ref{sec_approx}.

\begin{algorithm}
	\paper{
	\scriptsize
	}
	\caption{Encoding Algorithm}
	\label{alg_encode}
	\begin{algorithmic}
		\Function{ArithmeticCoding}{$[l_1,r_1], \ldots, [l_n,r_n]$}
			\State $\text{code} \leftarrow \emptyset$
			\State $I_t \leftarrow [0, 1]$
			\For{$i = 1$ to $n$}
				\State $I_t \leftarrow I_t \diamond [l_i, r_i]$
				\While{$\exists k=0 \text{ or } 1, I_t \subseteq [\frac{k}{2}, \frac{k+1}{2}]$}
					\State $\text{code} \leftarrow \text{code} + k$
					\State $I_t \leftarrow [2I_t.l - k, 2I_t.r - k]$
				\EndWhile
			\EndFor
			\State Find smallest $k$ such that \\ \qquad \qquad $\exists M, [2^{-k}M, 2^{-k}(M+1)] \subseteq I_t$
			\State \Return $\text{code} + M$
		\EndFunction
	\end{algorithmic}
\end{algorithm}

\vspace{-10pt}

\subsubsection{Early Bits Emission}\label{sec_early_bit_emission}

Without loss of generality, suppose
$$ [L,R] = [l_1, r_1] \circ [l_2, r_2] \circ \ldots \circ [l_n, r_n] $$
Define $[L_i,R_i]$ as the product of first $i$ probability intervals:
$$ [L_i,R_i] = [l_1, r_1] \circ [l_2, r_2] \circ \ldots \circ [l_i, r_i] $$
If there exist positive integer $k_i$ and non-negative integer $M_i$ such that
$$ 2^{-k_i} M_i \leq L_i < R_i \leq 2^{-k_i} (M_i + 1) $$
Then the first $k_i$ bits of the code string of $t$ must be the binary representation of $M_i$. Define 
$$[L'_i, R'_i] = [2^{k_i} L_i - M_i, 2^{k_i} R_i - M_i] $$
Then it can be verified that
\begin{align*}
& \textit{binary\_code}([L,R]) = \textit{binary\_code}(M_i) +\\
							   & \textit{binary\_code}([L'_i,R'_i] \circ [l_{i+1}, r_{i+1}] \ldots \circ [l_n,r_n])
\end{align*} 
Therefore, we can immediately output the first $k_i$ bits of the code string. After that, we compute the product:
$$ [L'_i,R'_i] \circ [l_{i+1}, r_{i+1}] \ldots \circ [l_n,r_n] $$
We can recursively use the same early bit emitting scheme for this product. In this way, we can greatly reduce the likelihood of precision overflow.

\subsubsection{Deterministic Approximation}\label{sec_approx}

For probability intervals containing $0.5$, we cannot emit any bits early. In rare cases, such a probability interval would exceed the precision limit, and the correctness of our algorithm would be compromised. 

To address this problem, we introduce the deterministic approximation trick.
Recall that the correctness of arithmetic coding relies on the non-overlapping property of the probability intervals. Therefore, we do not need to compute probability intervals with perfect accuracy: the correctness is guaranteed as long as we ensure these probability intervals do not overlap with each other.

Formally, let $t_1, t_2$ be two different tuples, and suppose their probability intervals are:
\begin{align*}
\text{PI}(t_1) = [l_1, r_1] = [l_{11}, r_{11}] \circ [l_{12}, r_{12}] \circ \ldots \circ [l_{1n_1}, r_{1n_1}]\\
\text{PI}(t_2) = [l_2, r_2] = [l_{21}, r_{21}] \circ [l_{22}, r_{22}] \circ \ldots \circ [l_{2n_2}, r_{2n_2}]
\end{align*}

The deterministic approximation trick is to replace $\circ$ operator with a deterministic operator $\diamond$ that approximates $\circ$ and has the following properties:
\begin{itemize}
	\item
	For any two probability intervals $[a,b]$ and $[c,d]$:
	$$ [a,b] \diamond [c,d] \subseteq [a,b] \circ [c,d] $$
	\item
	For any two probability intervals $[a,b]$ and $[c,d]$ with $b - a \geq \epsilon$ and $d - c \geq \epsilon$. Let $[l,r] = [a,b] \diamond [c,d]$, then:
	$$ \exists k, M, 2^{-k} M \leq l < r \leq 2^{-k} (M+1), 2^k(r - l) \geq \epsilon $$
\end{itemize}

In other words, the product computed by $\diamond$ operator is always a subset of the product computed by $\circ$ operator, and $\diamond$ operator always ensures that the product probability interval has length greater than or equal to $\epsilon$ after emitting bits. The first property guarantees the non-overlapping property still holds, and the second property prevents potential precision overflow. As we will see in Section~\ref{sec_decompression}, these two properties are sufficient to guarantee the correctness of arithmetic coding.

\techreport{
\subsection{Delta Coding}\label{sec_delta}
Notice that our compression scheme thus far has focused on compressing ``horizontally'', i.e., reducing the size of each tuple, independent of each other.
In addition to this, we could also compress ``vertically'', where
we compress tuples relative to each other. For this, we directly leverage an algorithm developed in prior work.
Raman and Swart~\cite{raman2006wring} developed an optimal method for compressing a set of binary code strings. This coding scheme (called ``Delta Coding'' in their paper) achieves $O(n \log n)$ space saving where $n$ is the number of code strings. For completeness, the pseudo-code of a variant of their method that is used in our system is listed in Algorithm~\ref{alg_delta}.

\begin{algorithm}
	\caption{Delta Coding}
	\label{alg_delta}
	\begin{algorithmic}
		\Function{DeltaCoding}{$s_1, \ldots, s_n$}
			\State // $s_1, \ldots, s_n$ are binary codes of $t_1, \ldots, t_n$
			\State Sort $s_1, \ldots, s_n$
			\State $l \leftarrow \lfloor \log n \rfloor$
			\State If $len(s_i) < l$, pad $s_i$ with trailing zeros
			\State Let $s_i = a_i b_i$ where $a_i$ is $l$-bit prefix of $s_i$
			\State $s'_i \leftarrow Unary(a_i - a_{i-1}) + b_i$
			\State \Return $\{s'_1, \ldots, s'_n\}$
		\EndFunction
	\end{algorithmic}
\end{algorithm}

In Algorithm~\ref{alg_delta}, $Unary(s)$ is the unary code of $s$ (i.e., $0 \rightarrow 0$, $1 \rightarrow 10$, $2 \rightarrow 110$, etc.). Delta Coding replaces the $\lfloor \log n \rfloor$-bit prefix of each tuple by an unary code with at most $2$ bits on average. Thus it saves about $n (\log_2 n - 2)$ bits storage space in total.
}

\subsection{Decompression}\label{sec_decompression}

When decompressing, \system~first reads in the dataset schema and all of the model information, and stores them in the main memory. After that, it scans over the compressed dataset, extracts and decodes the binary code strings to recover the original tuples.

\begin{algorithm}
	\paper{
	\scriptsize
	}
	\caption{Decoding Algorithm}
	\label{alg_decode}
	\begin{algorithmic}
		\Function{Decoder.Initialization}{}
			\State $I_b \leftarrow [0, 1]$
			\State $I_t \leftarrow [0, 1]$
		\EndFunction
		\Function{Decoder.GetNextBranch}{$branches$}
			\While{\textbf{not} $\exists br \in branches, I_b \subseteq I_t \diamond \text{PI}(br)$}
				\State Read in the next bit $x$
				\State $I_b \leftarrow I_b \circ [\frac{x}{2}, \frac{x+1}{2}]$
			\EndWhile
		\If{$I_b \subseteq I_t \diamond \text{PI}(br), br \in branches$}
			\State $I_t \leftarrow I_t \diamond \text{PI}(br)$
			\While{$\exists k=0 \text{ or } 1, I_t \subseteq [\frac{k}{2}, \frac{k+1}{2}]$}
				\State $I_t \leftarrow [2I_t.l - k, 2I_t.r - k]$
				\State $I_b \leftarrow [2I_b.r - k, 2I_b.r - k]$
			\EndWhile
			\State \Return $br$
		\EndIf
		\EndFunction
	\end{algorithmic}
\end{algorithm}

Algorithm~\ref{alg_decode} describes the procedure to decide the next branch. \techreport{Combined with decompression function in Algorithm~\ref{alg_api_calls}, we get the complete algorithm for decoding. 

}The decoder maintains two probability intervals $I_b$ and $I_t$. $I_b$ is the probability interval corresponding to all the bits that the algorithm has read in so far. $I_t$ is the probability interval corresponding to all decoded attributes. At each step, the algorithm computes the product of $I_t$ and the probability interval for every possible attribute value, and then checks whether $I_b$ is contained by one of those probability intervals. If so, we can decide the next branch, and update $I_t$ accordingly. If not, we continue reading in the next bit and update $I_b$.

\paper{By calling Algorithm~\ref{alg_decode} repeatedly, we can gradually decode the whole tuple. The full decoding procedure with an illustrative example can be found in our technical report~\cite{techreport}.}

\techreport{
As an illustration of the behavior of Algorithm~\ref{alg_decode}, Table~\ref{tbl_decode_example} shows the step by step execution for the following example:
$$ t = (a_1, a_2, a_3), [l,r] = [\frac{1}{3},\frac{1}{2}] \circ [\frac{1}{4}, \frac{1}{2}] \circ [\frac{1}{2}, \frac{2}{3}], \text{code} = 01100110 $$
\begin{table}[h]
\small
\vspace{-5pt}
	\centering
	\begin{tabular}{|c|c|c|c|c|}
		\hline
		Step & $I_t$ & $I_b$ & Input & Output \\
		\hline
		1 & $[0,1]$ & $[0,1]$ & & \\
		2 & $[0,1]$ & $[0,\frac{1}{2}]$ & 0 &\\
		3 & $[0,1]$ & $[\frac{1}{4},\frac{1}{2}]$ & 01 &\\
		4 & $[\frac{1}{3},\frac{1}{2}]$ & $[\frac{1}{4},\frac{1}{2}]$ & 01 & $a_1$ \\
		5 & $[\frac{2}{3},1]$ & $[\frac{1}{2},1]$ & 01 & $a_1$ \\
		6 & $[\frac{1}{3},1]$ & $[0,1]$ & 01 & $a_1$ \\	
		7 & $[\frac{1}{3},1]$ & $[\frac{1}{2},1]$ & 011 & $a_1$ \\
		8 & $[\frac{1}{3},1]$ & $[\frac{1}{2},\frac{3}{4}]$ & 0110 & $a_1$ \\
		9 & $[\frac{1}{3},1]$ & $[\frac{1}{2},\frac{5}{8}]$ & 01100 & $a_1$ \\
		10 & $[\frac{1}{2},\frac{2}{3}]$ & $[\frac{1}{2},\frac{5}{8}]$ & 01100 & $a_1,a_2$ \\
		11 & $[0,\frac{1}{3}]$ & $[0,\frac{1}{4}]$ & 01100 & $a_1,a_2$ \\
		12 & $[0,\frac{2}{3}]$ & $[0,\frac{1}{2}]$ & 01100 & $a_1,a_2$ \\
		13 & $[0,\frac{2}{3}]$ & $[\frac{1}{4},\frac{1}{2}]$ & 011001 & $a_1,a_2$ \\
		14 & $[0,\frac{2}{3}]$ & $[\frac{3}{8},\frac{1}{2}]$ & 0110011 & $a_1,a_2$ \\
		15 & $[0,\frac{2}{3}]$ & $[\frac{3}{8},\frac{7}{16}]$ & 01100110 & $a_1,a_2$ \\
		16 & $[\frac{1}{3},\frac{4}{9}]$ & $[\frac{3}{8},\frac{7}{16}]$ & 01100110 & $a_1,a_2,a_3$ \\
		17 & $[\frac{2}{3},\frac{8}{9}]$ & $[\frac{3}{4},\frac{7}{8}]$ & 01100110 & $a_1,a_2,a_3$ \\
		18 & $[\frac{1}{3},\frac{7}{9}]$ & $[\frac{1}{2},\frac{3}{4}]$ & 01100110 & $a_1,a_2,a_3$ \\
		\hline
	\end{tabular}
	\caption{Decoding Example}\label{tbl_decode_example}
	\vspace{-5pt}
\end{table}
}

Notice that Algorithm~\ref{alg_decode} mirrors Algorithm~\ref{alg_encode} in the way it computes probability interval products. This design is to ensure that the encoding and decoding algorithm always apply the same deterministic approximation that we described in Section~\ref{sec_approx}. The following theorem states the correctness of the algorithm (the proof can be found in \techreport{the appendix}\paper{our technical report~\cite{techreport}}):

\begin{theorem}\label{thm_decode}
	Let $[l_1,r_1],\ldots,[l_n,r_n]$ be probability intervals with $r_i - l_i \geq \epsilon$ where $\epsilon$ is the small constant defined in Section~\ref{sec_approx}. Let $s$ be the output of Algorithm~\ref{alg_encode} on these probability intervals.
	Then Algorithm~\ref{alg_decode} can always determine the correct branch from alternatives using $s$ as input:
	$$ \text{PI}(\text{Decoder.GetNextBranch}(branch_i)) = [l_i, r_i] $$
\end{theorem}

\techreport{
\section{Discussion: Examples and Optimality}\label{sec_optimality}

In this section, we use three example types of datasets to illustrate
how our compression algorithm can effectively find compact representations.
These examples also demonstrates the wide applicability of the \system\ approach.
We then describe the overall optimality of our algorithm.

\subsection{Illustrative Examples}\label{sec_example_dataset}

We now describe three types of datasets in turn.

\vspace{2mm}
\noindent \textbf{Pairwise Dependent Attributes}
\vspace{2mm}

Consider a dataset with 100 binary attributes $a_1, \ldots,$ $a_{100}$, where $a_1, \ldots, a_{50}$ are independent and uniformly distributed, and $a_{51}, \ldots,$ $ a_{100}$ are identical copies of $a_1, \ldots, a_{50}$ (i.e., $a_{i+50} = a_i$). 

Let us consider a tuple $(x_1, x_2, \ldots, x_{100})$. The probability intervals for the first $50$ attributes are $[\frac{x_i}{2}, \frac{x_i + 1}{2}]$ (i.e., $[0, \frac{1}{2}]$ if $x_i = 0$ and $[\frac{1}{2}, 1]$ otherwise). For the last $50$ attributes, since they deterministically depend on the first $50$ attributes, the probability interval is always $[0, 1]$. Therefore the probability interval for the whole tuple is:
$$ [\frac{x_1}{2}, \frac{x_1 + 1}{2}] \circ \ldots \circ [\frac{x_{50}}{2}, \frac{x_{50} + 1}{2}] \circ [0,1] \circ \ldots \circ [0,1] $$
It is easy to verify that the binary code string of the tuple is exactly the $50$-bits binary string $x_1 x_2\ldots x_{50}$ (recall that each $x_i$ is either $0$ or $1$). 

We also need to store the model information, which consists of the probability distribution of $x_1, \ldots, x_{50}$, the parent node of $x_{51}, \ldots, x_{100}$ and the conditional probability distribution of $x_{i+50}$ conditioned on $x_i$.

Assuming all the model parameters are stored using an $8$-bit code, then the model can be described using $200 \times 8 = 1600$ bits. Since each tuple uses $50$ bits, in total \system~would use $1600 + 50n$ bits for the whole dataset, where $n$ is the number of tuples. Note that our compression algorithm achieves much better compression rate than Huffman Coding~\cite{huffman1952method}, which uses at least $100$ bits per tuple.

Dependent attributes exist in many datasets. While they are usually only softly dependent (i.e., one attribute influences but does not completely determine the other attribute), our algorithm can still exploit these dependencies in compression.

\vspace{2mm}
\noindent \textbf{Markov Chain}
\vspace{2mm}

Figure~\ref{fig_markov} shows an example dataset with 1000 categorical attributes, in which each attribute $a_i$ depends on the preceding attribute $a_{i-1}$, and $a_1$ is uniformly distributed. This kind of dependency is called a {\em Markov Chain}, and frequently occurs in time-series datasets.

\begin{figure}[h]
	\centering
	\vspace{-5pt}	
	\includegraphics[width = 6cm]{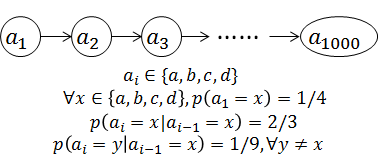}
	\vspace{-5pt}	
	\caption{Markov Chain Example}\label{fig_markov}
	\vspace{-5pt}	
\end{figure}

The probability interval of tuple $t = (x_1, \ldots, x_{1000})$ is:
$$ [\frac{x_1}{4}, \frac{x_1 + 1}{4}] \circ g(x_1, x_2) \circ g(x_2, x_3) \circ \ldots \circ g(x_{999}, x_{1000}) $$
where mapping $g(x,y)$ is listed below:
\begin{center}
	\begin{tabular}{|c|c|c|c|c|}
		\hline
		& y=1 & y=2 & y=3 & y=4 \\
		\hline
		x=1 & $[0,2/3]$ & $[2/3, 7/9]$ & $[7/9, 8/9]$ & $[8/9, 1]$ \\
		\hline
		x=2 & $[0,1/9]$ & $[1/9, 7/9]$ & $[7/9, 8/9]$ & $[8/9, 1]$ \\
		\hline
		x=3 & $[0,1/9]$ & $[1/9, 2/9]$ & $[2/9, 8/9]$ & $[8/9, 1]$ \\
		\hline
		x=4 & $[0,1/9]$ & $[1/9, 2/9]$ & $[2/9, 1/3]$ & $[1/3, 1]$ \\
		\hline
	\end{tabular}
\end{center}

On average, for each tuple our algorithm uses about
$$ 1000 \times (\frac{2}{3} \log_2 \frac{3}{2} + 3 \times \frac{1}{9} \log_2 9) \approx 1443 \text{ bits}$$
while standard Huffman Coding~\cite{huffman1952method} uses $2000$ bits.

Time series datasets usually contain a lot of redundancy that can be used to achieve significant compression. As an example, most electrocardiography (ECG) waveforms~\cite{jalaleddine1990ecg} can be restored using a little extra information if we know the cardiac cycle. Our algorithm offers effective ways to utilize such redundancies for compression.

\vspace{2mm}
\noindent \textbf{Clustered Tuples}
\vspace{2mm}

Figure~\ref{fig_cluster} shows an example with 100 binary attributes: $a_1, \ldots, a_{100}$. In this example, $c$ is the hidden cluster index attribute and all other attributes are dependent on it.

\begin{figure}[h]
	\centering
	\vspace{-5pt}	
	\includegraphics[width = 5cm]{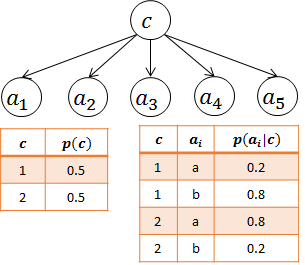}
	\vspace{-5pt}	
	\caption{Clustered Tuples Example}\label{fig_cluster}
	\vspace{-5pt}	
\end{figure}

If we compress the cluster index together with all attributes using our algorithm, we will need about
$$ H(t_i) = 1 + 100 \times (0.2 \log_2 \frac{1}{0.2} + 0.8 \log_2 \frac{1}{0.8}) \approx 73\text{ bits} $$
for each tuple. Note that it is less than the $100$-bits used by the plain binary code.

Many real datasets have a clustering property. To compute the cluster index, we need to choose an existing clustering algorithm that is most appropriate to the dataset at hand~\cite{jain1999data}. The extra cost of storing a cluster index is usually small compared to the overall saving in compression.

\subsection{Asymptotic Optimality}
}
\paper{
\section{Discussion: Optimality}\label{sec_optimality}
}

We can prove that \system~achieves asymptotic near-optimal compression rate for lossless compression if the dataset only contains categorical attributes and can be described efficiently using a Bayesian network (the proof can be found in\paper{~\cite{techreport}}\techreport {the appendix}):

\begin{theorem}\label{thm_asymptotic}
	Let $a_1, a_2, \ldots, a_m$ be categorical attributes with joint probability distribution $P(a_1, \ldots, a_m)$ that decomposes as
	$$ P(a_1, \ldots, a_m) = \prod_{i=1}^m P(a_i|\text{parent}_i) $$
	such that
	$$ \text{parent}_i \subseteq \{a_1, \ldots, a_{i-1}\}, \text{card}(\text{parent}_i) \leq c $$
	
	Suppose the dataset $\mathcal{D}$ contains $n$ tuples that are i.i.d. samples from $P$. Let $M = \max_i \text{card}(a_i)$ be the maximum cardinality of attribute range. Then \system~can compress $\mathcal{D}$ using less than $H(\mathcal{D}) + 4n + 32 m M^{c+1}$ bits on average, where $H(\mathcal{D})$ is the entropy~\cite{cover2012elements} of the dataset $\mathcal{D}$.
\end{theorem}

Thus, when $n$ is large, the difference between the size of the compressed dataset using our system and the entropy\footnote{By  Shannon's source coding theorem~\cite{cover2012elements}, there is no algorithm that can achieve compression rate higher than entropy.} of $\mathcal{D}$ is at most $5n$, that is only $5$ bits per tuple. This indicates that \system~is asymptotically near-optimal for this setting.

When the dataset $\mathcal{D}$ contains numerical attributes, the entropy $H(\mathcal{D})$ is not defined, and the techniques we used to prove Theorem~\ref{thm_asymptotic} no longer apply. However, in light of Theorem~\ref{thm_numerical}, it is likely that \system~still achieves asymptotic near-optimal compression. 


\section{Experiments}\label{sec_experiment}

In this section, we evaluate the performance of \system~against the state of the art semantic compression algorithms SPARTAN~\cite{babu2001spartan} and ItCompress~\cite{jagadish2004itcompress} (see Table~\ref{tbl_intro}). For reference we also include the performance of gzip~\cite{ziv1977universal}, a well-known syntactic compression algorithm. 

We use the following four publicly available datasets:
\begin{itemize}
\item
\textit{Corel} (http://kdd.ics.uci.edu/databases/CorelFeatures) is a 20 MB dataset containing 68,040 tuples with 32 numerical color histogram attributes.
\item
\textit{Forest-Cover} (http://kdd.ics.uci.edu/databases/covertype) is a 75 MB dataset containing 581,000 tuples with $10$ numerical and $44$ categorical attributes.
\item
\textit{Census} (http://thedataweb.rm.census.gov/ftp/cps\_ftp.html) is a 610MB dataset containing 676,000 tuples with $36$ numerical and $332$ categorical attributes.
\item
\textit{Genomes} (ftp://ftp.1000genomes.ebi.ac.uk) is a 18.2GB dataset containing 1,832,506 tuples with about $10$ numerical and $2500$ categorical attributes.\footnote{In this dataset, many attributes are optional and these numbers indicate the average number of attributes that appear in each tuple.}
\end{itemize}

The first three datasets have been used in previous papers~\cite{babu2001spartan, jagadish2004itcompress}, and the compression ratio achieved by SPARTAN, ItCompress and gzip on these datasets have been reported in Jagadish et al.'s work~\cite{jagadish2004itcompress}. We did not reproduce these numbers and only used their reported performance numbers for comparison. For the \textit{Census} dataset, the previous papers only used a subset of the attributes in the experiments ($7$ categorical attributes and $7$ numerical attributes). Since we are unaware of the selection criteria of the attributes, we are unable to compare with their algorithms, and we will only report the comparison with gzip.

For the \textit{Corel} and \textit{Forest-Cover} datasets, we set the error tolerance as a percentage ($1\%$ by default) of the width of the range for numerical attributes as in previous work. For the \textit{Census} dataset, we set all error tolerance to $0$ (i.e. the compression is lossless). For the \textit{Genomes} dataset, we set the error tolerance for integer attributes to $0$ and float number attributes to $10^{-8}$.

In all experiments, we only used the first 2000 tuples in the structure learning algorithm \techreport{(Algorithm~\ref{alg_structure_learning}) }to improve efficiency. All available tuples are used in other parts of the algorithm.

\techreport{
	We remark that unlike experiments in machine learning, there is no training/test split of the dataset in our setting. We always use all the available data to train the model, and the goal is simply to minimize the size of the compressed dataset. There is no concept of overfitting in this scenario.
}

\subsection{Compression Rate Comparison}

\begin{figure}[h]
	\centering
	
	\subfigure[Forest Cover]{
		\includegraphics[width = 6cm]{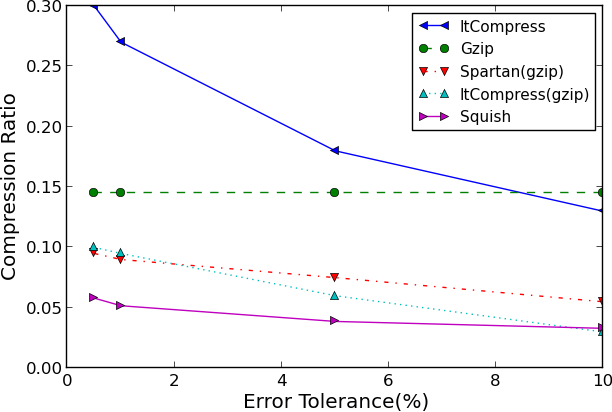}
		\vspace{-5pt}
		\label{fig_ratio_forest_cover}
	}
	\subfigure[Corel]{
		\includegraphics[width = 6cm]{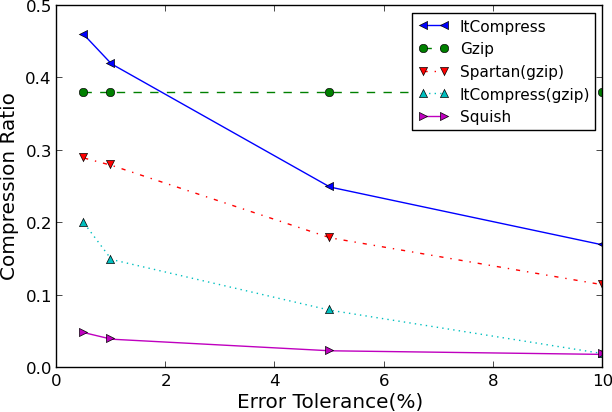}
		\vspace{-5pt}
		\label{fig_ratio_corel}
	}
	\vspace{-10pt}	
	\caption{Error Threshold vs Compression Ratio}\label{fig_ratio}
\end{figure}

Figure~\ref{fig_ratio} shows the comparison of compression rate on the \textit{Corel} and \textit{Forest-Cover} datasets. In these figures, X axis is the error tolerance for numerical attributes (\% of the width of range), and Y axis is the compression ratio, defined as follows:
$$ \text{compression ratio} = \frac{\text{data size with compression}}{\text{data size without compression}} $$

As we can see from the figures, \system~significantly outperforms the other algorithms. When the error tolerance threshold is small (0.5\%), \system~achieves about {\em 50\% reduction in compression ratio on the Forest Cover dataset and 75\% reduction on the Corel dataset, compared to the nearest competitor ItCompress (gzip)}, which applies gzip algorithm on top of the result of ItCompress. The benefit of not using gzip as a post-processing step is that we can still permit tuple-level access without decompressing a larger unit.

The remarkable superiority of our system in the Corel dataset reflects the advantage of \system~in compressing numerical attributes. Numerical attribute compression is known to be a hard problem~\cite{roth1993database} and none of the previous systems have effectively addressed it. In contrast, our encoding scheme can leverage the skewness of the distribution and achieve near-optimal performance.

\begin{figure}[h]
	\centering
	\includegraphics[width = 6cm]{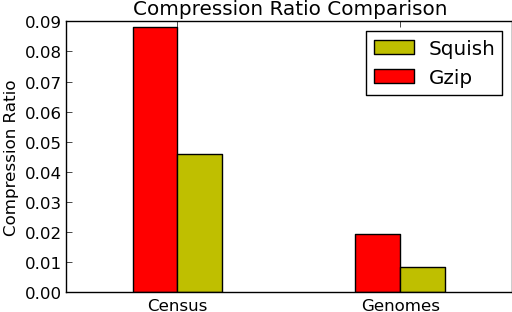}
	\vspace{-10pt}	
	\caption{Compression Ratio Comparison}\label{fig_ratio_lossless}
	\vspace{-10pt}	
\end{figure}

Figure~\ref{fig_ratio_lossless} shows the comparison of compression rate on the \textit{Census} and \textit{Genomes} datasets. Note that in these two datasets, we set the error tolerance threshold to be extremely small, so that the compression is essentially lossless. As we can see, even in the lossless compression scenario, our algorithm still outperforms gzip significantly. {\em Compared to gzip, \system~achieves $48\%$ reduction in compression ratio in Census dataset and $56\%$ reduction in Genomes dataset.}

\subsection{Compression Breakdown}

As we have seen in the last section, \system~achieved superior compression ratio in all four datasets. In this section, we use detailed case studies to illustrate the reason behind the significant improvement over previous papers.

\subsubsection{Categorical Attributes}

In this section, we study the source of the compression in \system~for categorical attributes. We will use three different treatments for the categorical attributes and see how much compression is achieved for each of these treatments: 
\begin{itemize} \setlength{\itemsep}{0ex} \setlength{\topsep}{0ex}
	\item
	Domain Code: We replace the categorical attribute values with short binary code strings. Each code string has length $\lceil \log_2 N \rceil$, where $N$ is the total number of possible categorical values for the attribute. 
	\item
	Column: We ignore the correlations between categorical attributes and treat all the categorical attributes as independent. 
	\item
	Full: We use both the correlations between attributes and the skewness of attribute values in our compression algorithm.
\end{itemize}

We will use the \textit{Genomes} and \textit{Census} dataset here since they consist of mostly categorical attributes. We keep the compression algorithm for numerical attributes unchanged in all treatments. Figure~\ref{fig_ratio_categorical} shows the compression ratio of the three treatments:

\begin{figure}[h]
	\centering
	\vspace{-10pt}	
	\includegraphics[width = 5cm]{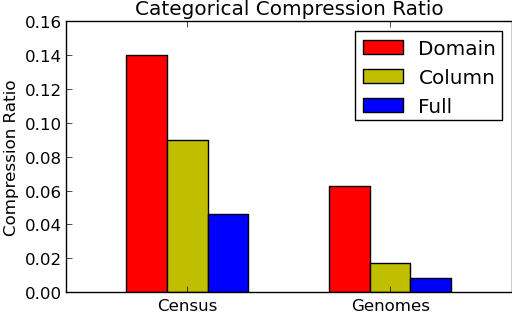}
	\vspace{-10pt}	
	\caption{Compression Ratio Comparison}\label{fig_ratio_categorical}
	\vspace{-10pt}	
\end{figure}

As we can see, the compression ratio of the basic domain coding scheme can be improved up to $70$\% if we take into account the skewness of the distribution in attribute values. Furthermore, the correlation between attributes is another opportunity for compression, which improved the compression ratio by $50\%$ in both datasets.

An interesting observation is that the Column treatment achieves comparable compression ratio as gzip in both datasets, which suggests that gzip is in general capable of capturing the skewness of distribution for categorical attributes, but unable to capture the correlation between attributes.

\subsubsection{Numerical Attributes}

We now study the source of the compression in \system \\for numerical attributes. We use the following five treatments for the numerical attributes:
\begin{itemize} \setlength{\itemsep}{0ex} \setlength{\topsep}{0ex}
	\item
		IEEE Float: We use the IEEE Single Precision Floating Point standard to store all attributes.
	\item
		Discrete: Since all attributes in the dataset have value between $0$ and $1$, we use integer $i$ to represent a float number in range $[\frac{i}{10^7}, \frac{i + 1}{10^7}]$, and then store each integer using its 24-bit binary representation.
	\item
		Column: We ignore the correlation between numerical attributes and treat all attributes as independent.
	\item
		Full: We use both the correlations between attributes and distribution information about attribute values.
	\item
		Lossy: The same as the Full treatment, but we set the error tolerance at $10^{-4}$ instead.
\end{itemize}

\begin{figure}[h]
	\centering
	\vspace{-5pt}	
	\includegraphics[width = 5cm]{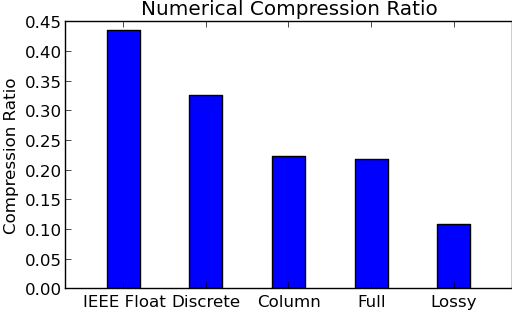}
	\vspace{-10pt}	
	\caption{Compression Ratio Comparison}\label{fig_ratio_numerical}
	\vspace{-5pt}	
\end{figure}

We use the \textit{Corel} dataset here since it contains only numerical attributes. The error tolerance in all treatments except the last are set to be $10^{-7}$ to make sure the comparison is fair (IEEE single format has precision about $10^{-7}$). All the numerical attributes in this dataset are in range $[0, 1]$, with a distribution peaked at $0$. Figure~\ref{fig_ratio_numerical} shows the compression ratio of the five treatments.

As we can see, storing numerical attributes as float numbers instead of strings gives us about $55\%$ compression. However, the compression rate can be improved by another $50\%$ if we recognize distributional properties (i.e., range and skewness). Utilizing the correlation between attributes in the \textit{Corel} dataset only slightly improved the compression ratio by $3\%$. Finally, we see that the benefit of lossy compression is significant: even though we only reduced the precision from $10^{-7}$ to $10^{-4}$, the compression ratio has already been improved by $50\%$.

\subsection{Running Time}

\techreport{In this section we evaluate the running time of \system. Note that the time complexity of the compression and decompression components are both $O(n m)$, where $m$ is the number of attributes and $n$ is the number of tuples. In other words, the running time of these two components are linear to the size of the dataset. Therefore, the algorithm should scale well to large datasets in theory.	
}

Table~\ref{tbl_running_time} lists the running time of the five components in \system. All experiments are performed on a computer with eight\footnote{The implementation is single-threaded, so only one processor is used.} 3.4GHz Intel Xeon processors. For the \textit{Genomes} dataset, which contains $2500$ attributes---an extremely large number---we constructed the Bayesian Network manually. 
Note that none of the previous papers have been applied on a dataset with the magnitude of the \textit{Genomes} dataset (both in number of tuples and number of attributes). 

\techreport{
\textbf{Remark:} Due to a suboptimal implementation of the parameter tuning component in our current code, the actual time complexity of the parameter tuning component is $O(n m d)$ where $d$ is the depth of the Bayesian network. Therefore, Table~\ref{tbl_running_time} may not reflect the best possible running time of a fully optimized version of our compression algorithm.	
}
\begin{table}[h]
	\vspace{-15pt}
	\caption{Running Time of Different Components}\label{tbl_running_time}
	\centering
	\small
	\begin{tabular}{|c|c|c|c|c|}
		\hline
		 & Forest Cov. & Corel & Census & Genomes\\
		\hline
		Struct. Learning & 5.5 sec & 2.5 sec & 20 min & N/A \\
		\hline
		Param. Tuning & 140 sec & 15 sec & 100 min & 40 min\\
		\hline
		Compression & 48 sec & 6 sec & 6 min & 50 min\\
		\hline
		Writing to File & 7 sec & 2 sec & 40 sec & 7 min\\
		\hline
		Decompression & 53 sec & 7.5 sec & 6 min & 50 min\\
		\hline
	\end{tabular}
	\vspace{-10pt}
\end{table}


As we can see from Table~\ref{tbl_running_time}, our compression algorithm scales reasonably: even with the largest dataset \textit{Genomes}, the compression can still be finished within hours. Recall that since our algorithm is designed for archival not online query processing, and {\em our goal is therefore to minimize storage as much as possible}, a few hours for large datasets is adequate.

The running time of the parameter tuning component can be greatly reduced if we use only a subset of tuples (as we did for structure learning). The only potential bottleneck is structure learning, which scales badly with respect to the number of attributes ($O(m^4)$). To handle datasets of this scale, another approach is to partition the dataset column-wise, and apply the compression algorithm on each partition separately. We plan to investigate this in future work.


We remark that, unlike gzip~\cite{ziv1977universal}, \system~allows random access of tuples without decompressing the whole dataset. Therefore, if users only need to access a few tuples in the dataset, then they will only need to decode those tuples, which would require far less time than decoding the whole dataset.

\subsection{Sensitivity to Bayesian Network Learning}

We now investigate the sensitivity of the performance of our algorithm with respect to the Bayesian network learning. We use the \textit{Census} dataset here since the correlation between attributes in this dataset is stronger than other datasets, so the quality of the Bayesian network can be directly reflected in the compression ratio.

Since our structure learning algorithm only uses a subset of the training data, one might question whether the selection of tuples in the structure learning component would affect the compression ratio. To test this, we run the algorithm for five times, and randomly choose the tuples participating in the structure learning. Table~\ref{tbl_sensitivity} shows the compression ratio of the five runs. As we can see, the variation between runs are insignificant, suggesting that our compression algorithm is robust.

\begin{table}[h]
	\vspace{-10pt}
	\caption{Sensitivity of the Structure Learning}\label{tbl_sensitivity}
	\centering
	\begin{tabular}{|c|c|c|c|c|c|}
		\hline
		No. of Exp. & 1 & 2 & 3 & 4 & 5\\
		\hline
		Comp. Ratio & 0.0460 & 0.0472 & 0.0471 & 0.0468  & 0.0476 \\
		\hline
	\end{tabular}
	\vspace{-5pt}
\end{table}

We also study the sensitivity of our algorithm with respect to the number of tuples used for structure learning. Table~\ref{tbl_tuples} shows the compression ratio when we use $1000$, $2000$ and $5000$ tuples in the structure learning algorithm respectively. As we can see, the compression ratio improves gradually as we use more tuples for structure learning.

\begin{table}[h]
	\vspace{-15pt}
	\caption{Sensitivity to Number of Tuples}\label{tbl_tuples}
	\centering
	\begin{tabular}{|c|c|c|c|}
		\hline
		Number of Tuples & 1000 & 2000 & 5000 \\
		\hline
		Comp. Ratio & 0.0474 & 0.0460 & 0.0427 \\
		\hline
	\end{tabular}
	\vspace{-15pt}
\end{table}

\section{Related Work}\label{sec_related_work}

Although compression of datasets is a classical research topic in the database research community~\cite{roth1993database}, the idea of exploiting attribute correlations (a.k.a. semantic compression) is relatively new. Babu et al.~\cite{babu2001spartan} used functional dependencies among attributes to avoid storing them explicitly. Jagadish et al.~\cite{jagadish2004itcompress} used a clustering algorithm for tuples. Their compression scheme stores, for each cluster of tuples, a representative tuple and the differences between the representative tuple and other tuples in the cluster. These two types of dependencies are special cases of the more general Bayesian network style dependencies used in this paper.

The idea of applying arithmetic coding on Bayesian networks was first proposed by Davies and Moore~\cite{davies1999bayesian}. However, their work only supports categorical attributes (a simple case). Further, the authors did not justify their approach by either theoretically or experimentally comparing their algorithm with other semantic compression algorithms. Lastly, they used conventional BIC~\cite{schwarz1978estimating} score for learning a Bayesian Network, which is suboptimal, and their technique does not apply to the lossy setting.


The compression algorithm developed by Raman and Swart~\cite{raman2006wring} used Huffman Coding to compress attributes. Therefore, their work can only be applied to categorical attributes and can not fully utilize attribute correlation (the authors only mentioned that they can exploit attribute correlations by encoding multiple attributes at once). The major contribution of Raman's work~\cite{raman2006wring} is that they formalized the old idea of compressing ordered sequences by storing the difference between adjacent elements, which has been used in search engines to compress inverted indexes~\cite{baeza1999modern} and also in column-oriented database systems~\cite{stonebraker2005c}. 

Bayesian networks are well-known general purpose probabilistic models to characterize dependencies between random variables. For reference, the textbook written by Koller and Friedman~\cite{koller2009probabilistic} covers many recent developments. Arithmetic coding was first introduced by Rissanen~\cite{rissanen1976generalized} and further developed by Witten et al.~\cite{witten1987arithmetic}. An introductory paper written by Langdon Jr.~\cite{langdon1984introduction} covers most of the basic concepts of Arithmetic Coding, including the early bit emission trick. The deterministic approximation trick is original. Compared to the overflow prevention mechanism in Witten et al.'s work~\cite{witten1987arithmetic}, the deterministic approximation trick is simpler and easier to implement.


\section{Conclusion}\label{sec_conclusion}

In this paper, we propose \system, an extensible system for compressing relational datasets. \system~exploits both correlations between attributes and skewness of numerical attributes, and thereby achieves better compression rates than prior work. We also develop \probtree, an interface for supporting user-defined attributes in \system. Users can use \probtree~to define new data types by simply implementing a handful of functions. We develop new encoding schemes for numerical attributes using \probtree, and prove its optimality. We also discuss mechanisms for ensuring the correctness of Arithmetic Coding in finite precision systems. We prove the asymptotic optimality of \system~on any dataset that can be efficiently described using a Bayesian Network. Experiment results on two real datasets indicate that \system~significantly outperforms prior work, achieving more than 50\% reduction in storage. 

\section*{Acknowledgement}
We thank the anonymous reviewers for their valuable feedback. We acknowledge support from grant IIS-1513407 awarded by the National Science Foundation, grant 1U54GM114838 awarded by NIGMS and 3U54EB020406-02S1 awarded by NIBIB through funds provided by the trans-NIH Big Data to Knowledge (BD2K) initiative (www.bd2k.nih.gov), the Siebel Energy Institute, and the Faculty Research Award provided by Google. The content is solely the responsibility of the authors and does not necessarily represent the official views of the funding agencies and organizations.

\balance
{\small
\bibliographystyle{abbrv}
\bibliography{dbbib}
}
\techreport{
\appendix{

\subsection*{Proof of Theorem~\ref{thm_numerical}}

\vspace{5pt}

\begin{proof}
	For any $x \in \mathcal{X}$, define $S(x)$ as follows:
	$$ S(x) = \{y \in \mathcal{X} : g(y) = g(x)\} $$
	Then the probability that $g(x)$ is the code word is:
	$$ \mathbf{Pr}(g(x)) = \int_{y \in S(x)} f(y) dy $$
	The entropy of $g(x)$ is:
	$$ H(g(x)) = E_X[- \log_2 \mathbf{Pr}(g(X))] $$
	Since $g(x)$ is uniquely decodable, the average length must be greater than or equal to the entropy:
	$$ E_X[len(g(X))] \geq H(g(x)) $$
	For the first part, since $- \log x$ is decreasing function, it suffices to prove that
	$$ \forall x \in \mathcal{X}, \mathbf{Pr}(g(x)) \leq (2 \rho(x) \epsilon + 1) 4 \epsilon f(x) $$
	Note that due to closeness constraint, we have
	$$ S(x) \subseteq [h(g(x)) - \epsilon, h(g(x)) + \epsilon]$$
	Combined with the fact that $|h(g(x)) - x| \leq \epsilon$, we have
	$$ S(x) \subseteq [x - 2\epsilon, x + 2\epsilon] $$
	and
	$$ \mathbf{Pr}(g(x)) = \int_{y \in S(x)} f(y) dy \leq \int_{x - 2\epsilon}^{x + 2\epsilon} f(y) dy $$
	The right hand side can be upper bounded by
	$$ \int_{x - 2\epsilon}^{x + 2\epsilon} f(y) dy \leq 4 \epsilon (2 \epsilon \rho(x) + 1) f(x) $$
	since by Equation~(\ref{eqn_self_bounded}),
	$$\forall y \in \mathcal{X}, |x - y| < 2\epsilon \Rightarrow f(y) \leq (2 \epsilon \rho(x) + 1) f(x)$$
	
	For the second part, by the property of Arithmetic Coding, we have
	$$ E_X[len(g(X))] \leq H(g(x)) + 2 $$
	We will prove
	\begin{equation}\label{eqn_aux}
	\forall x \in \mathcal{X}, \mathbf{Pr}(g(x)) \geq \min(\frac{l}{2}, \frac{1}{4 \rho(x)}) f(x)
	\end{equation}
	Let us suppose Equation~(\ref{eqn_aux}) is true for now, then we have
	\begin{align*}
	H(g(X)) = & E_X[- \log_2 \mathbf{Pr}(g(x))] \\
	\leq & E_X[- \log_2 \min(\frac{l}{2}, \frac{1}{4 \rho(x)}) f(x)] \\
	= & E_X[- \log_2 f(x)] - \\
	& \quad E_X[\log_2 l + \min(\log_2 \frac{1}{2 \rho(x)} - \log_2 l, 0)] + 1 \\
	\leq & E_X[- \log_2 f(x)] - \log_2 l + \\
	& \quad E_X[\max(\log_2 \rho(x) + \log_2 l, 0)] + 2
	\end{align*}
	
	To prove Equation~(\ref{eqn_aux}), let $v$ be the leaf node that $x$ corresponds to, we consider two cases:
	\begin{enumerate}
		\item
		If $v_r - v_l < \frac{1}{2 \rho(x)}$, then
		$$ \forall y \in [v_l, v_r], f(y) \geq (1 - (v_r - v_l) \rho(x)) f(x) \geq \frac{1}{2} f(x) $$
		thus we have
		$$ \mathbf{Pr}(g(x)) \geq \frac{1}{2} (v_r - v_l) f(x) \geq \frac{1}{2} l f(x) $$
		\item
		If $v_r - v_l \geq \frac{1}{2 \rho(x)}$, then
		$$ \forall y \in [x - \frac{1}{2 \rho(x)}, x + \frac{1}{2 \rho(x)}] \cap [v_l, v_r], f(y) \geq \frac{1}{2} f(x) $$
		Therefore,
		$$ \mathbf{Pr}(g(x)) \geq \frac{1}{2} f(x) len([v_l,v_r] \cap [x - \frac{1}{2 \rho(x)}, x + \frac{1}{2 \rho(x)}]) $$
		Since $x \in [v_l, v_r]$, $[x - \frac{1}{2 \rho(x)}, x + \frac{1}{2 \rho(x)}]$ covers at least $\frac{1}{2 \rho(x)}$ length of $[v_l, v_r]$, therefore,
		$$ \mathbf{Pr}(g(x)) \geq \frac{1}{4 \rho(x)} f(x)  $$
	\end{enumerate}
	Since $\mathbf{Pr}(g(x))$ is always greater than or equal to at least one of terms, it must always be greater than or equal to the minimum of two. Thus Equation~(\ref{eqn_aux}) is proved.
\end{proof}

\subsection*{Proof of Theorem~\ref{thm_decode}}

\vspace{5pt}

\begin{proof}
	First we briefly examine Algorithm~\ref{alg_encode}. Define $a_i, b_i, c_i$ as follows:
	\begin{align*} 
	& a_i = b_{i-1} \diamond [l_i, r_i] (b_0 = [0,1]) \\
	& b_i = [2^{k_i} a_i.l - M_i, 2^{k_i} a_i.r - M_i] \\
	& c_i = c_{i-1} \circ [2^{-k_i}M_i, 2^{-k_i}(M_i + 1)]
	\end{align*}
	where $k_i$ is the largest integer such that 
	$$a_i \subseteq [2^{-k_i} M_i, 2^{-k_i} (M_i + 1)]$$
	Then, we have the following results:
	\begin{itemize}
		\item
		$a_i$ is the value of $I_t$ after executing the first step of $i$th iteration of the loop, $b_i$ is the value of $I_t$ at the end of $i$th iteration of the loop,
		and $c_i$ is the probability interval corresponding to $\text{code}$.
		\item
		$s$ is the binary code of the probability interval $c_n \circ b_n$.
		
		Define $\text{PI}_s = [2^{-len(s)}s, 2^{-len(s)}(s + 1)]$, then 
		$$\text{PI}_s \subseteq c_n \circ b_n \subseteq c_i \circ b_i = c_{i-1} \circ a_i$$
	\end{itemize}
	
	We can prove the following statements by induction on the steps of Algorithm~\ref{alg_decode} (see Table~\ref{tbl_decode_example} for reference):
	\begin{itemize}
		\item
		During the procedure of determining $i$th branch, the value of $I_t$ is:
		\begin{itemize}
			\item
			$b_{i-1}$, during the first while loop.
			\item
			$a_i$, after executing the first statement inside the if block.
			\item
			$b_i$, at the end of the if block.
		\end{itemize}
		\item
		During the procedure of determining $i$th branch, let $d$ be the input that the algorithm has read in. Define
		$$ \text{PI}_d = [2^{-len(d)}d, 2^{-len(d)}(d + 1)] $$
		Then the value of $I_b$ satisfies:
		\begin{itemize}
			\item
			$\text{PI}_d = c_{i-1} \circ I_b$, during the first while loop.
			\item
			$\text{PI}_d = c_i \circ I_b$, at the end of the if block.
		\end{itemize}
		\item
		$ I_b \subseteq I_t $ always holds.
	\end{itemize}
	The induction step is easy for most parts, we will only prove the nontrivial part that the loop condition is satisfied before reaching the end of input. i.e., after certain steps we must have
	$$ I_b \subseteq I_t \diamond [l_i, r_i] $$
	To prove this, note that
	\begin{align*}
	\text{PI}_s \subseteq \text{PI}_d = c_{i-1} \circ I_b \\
	\text{PI}_s \subseteq c_{i-1} \circ a_i
	\end{align*}
	Therefore,
	$$(c_{i-1} \circ I_b) \cap (c_{i-1} \circ a_i) \neq \emptyset \Leftrightarrow I_b \cap a_i \neq \emptyset$$
	Note that $a_i = b_{i-1} \diamond [l_i, r_i] = I_t \diamond [l_i, r_i]$ during the while loop. Thus for any other branch $[l',r']$, $I_b \not\subseteq I_t \diamond [l',r']$ due to the fact that $[l', r'] \cap [l_i, r_i] = \emptyset$. Also note that as we continue reading new bits, $\text{PI}_d = c_{i-1} \circ I_b$ approaches $\text{PI}_s$, thus eventually we would have:
	$$ c_{i-1} \circ I_b \subseteq c_{i-1} \circ a_i \Leftrightarrow I_b \subseteq a_i $$
\end{proof}

\subsection*{Proof of Theorem~\ref{thm_asymptotic}}

\vspace{5pt}

\begin{proof}
	Since our compression algorithm searches for the Bayesian Network with minimum description length, it suffices to prove that the size of compressed dataset using the correct Bayesian Network is less than $H(\mathcal{D}) + 4n + 32m M^{c+1}$.
	
	Let $S$ be the set of tuples in $\mathcal{D}$ such that its probability is less than $2^{-n}$:
	$$ S = \{i \in [n] : P(t_i) < 2^{-n}\} $$
	Then, the size of the compressed dataset can be expressed as:
	\begin{align*}
	\text{compressed size} \leq & \sum_{i \in S} (- \log_2 P(t_i) - (\log_2 n - 2) + 2) \\
	& + 2 (n - |S|) + (\text{cost of model description})\\
	\leq & \sum_{i \in S} (- \log_2 P(t_i)) - |S| \log_2 |S| \\
	& + 4n + 32 m M^{c+1}
	\end{align*}
	where we assume that the model parameters are stored using single precision float numbers.
	
	On the other hand, since $S$ deterministically depends on $\mathcal{D}$, therefore by the chain rule of entropy we have:
	\begin{align*}
	& H(\mathcal{D}) = H(\mathcal{D}) + H(S|\mathcal{D}) = H(\mathcal{D}, S) \\
	= & H(S) + H(t_S|S) + H(t_{[n] \setminus S}|S) \geq H(t_S|S)
	\end{align*}
	and since $t_S$ is a multi-set consisting of $\{t_i : i \in S\}$, we have
	$$ H(t_S|S) \geq - |S| E[\log_2 P(t) | P(t) < 2^{-n}] - |S| \log_2 |S| $$	
	
	Combining these two directions, we conclude that
	$$ \text{compressed size} \leq H(\mathcal{D}) + 4n + 32m M^{c+1} $$
\end{proof}

}
}
\end{document}